\documentclass[11pt,a4paper,english,superscriptaddress,aps,prd,preprint]{revtex4-1}
\usepackage[latin9]{inputenc}
\usepackage{babel}
\usepackage{amsmath}
\usepackage{amssymb}
\usepackage{graphicx}
\usepackage[unicode=true,
 bookmarks=false,
 breaklinks=false,pdfborder={0 0 1},backref=section,colorlinks=false]
 {hyperref}

\makeatletter

\pdfpageheight\paperheight
\pdfpagewidth\paperwidth

\usepackage{babel}
\usepackage{color}

\makeatother

\begin{document}

\title{$CP^{(N-1)}$ model in aether-superspace}

\author{A. F. Ferrari}
\email{alysson.ferrari@ufabc.edu.br}
\affiliation{\emph{Universidade Federal do ABC - UFABC, Rua Santa Adélia, 166,
09210-170, Santo André, SP, Brazil}}
\affiliation{Indiana University Center for Spacetime Symmetries, Indiana University,
Bloomington, Indiana 47405-7105}

\author{A. C. Lehum}
\email{lehum@ufpa.br}
\affiliation{Faculdade de Física, Universidade Federal do Pará, 66075-110, Belém,
Pará, Brazil}

\begin{abstract}
In this paper we study the dynamical generation of mass in the Lorentz-violating
$CP^{(N-1)}$ model defined in two and three-dimensional aether-superspace.
We show that even though the model presents a phase structure similar
to the usual, Lorentz invariant case, the dynamically generated mass
by quantum corrections has a dependence on the Lorentz violating background
properties, except for spacelike LV vector parameter. This is to be
contrasted with the behavior of the quantum electrodynamics in the
two-dimensional aether-superspace, where the dynamical generation
of mass was shown to exhibit an explicit dependence on the aether
parameters in every possible case. 
\end{abstract}


\maketitle

\section{Introduction}

The possibility of small deviations of the Lorentz symmetry have been
much studied in the last years, from the theoretical and the experimental
approach. The motivation is the idea that new physics at the Planck
scale might modify the symmetry or even the very nature of spacetime
in very small length scales. One might imagine a more fundamental
theory involving tensorial fields which acquire non vanishing vacuum
expectation values, appearing in low energy as the coupling of Standard
Model fields to constant background tensors: in this way, the Standard
Model could be understood as an effective field theory receiving small
Lorentz violating corrections. This approach was systematized in\,\cite{colladay:1998fq},
paving the way for a rich set of experimental investigations and,
as a consequence, very stringent constraints on the Lorentz violation
(LV) coefficients\,\cite{Kostelecky:2008ts}. 

The same idea can also be implemented in the context of supersymmetric
models. Supersymmetry, at the more fundamental level, is understood
as a generalization of Poincaré symmetry, yet one may incorporate
in the supersymmetric algebra translation invariant deformations of
spacetime, including Lorentz violation, since the supercharges and
the translation generators form a closed sub-algebra\,\cite{Terashima:2000xq}.
As an example, in the context of models with spacetime noncommutativity
extensively studied in the first decade of this century, where LV
appears naturally from the string background that originates the noncommutativity\,\cite{seiberg:1999vs},
supersymmetry was argued to be a natural way to avoid difficulties
in the perturbative consistency of such models\,\cite{girotti:2000gc,ferrari:2003vs,ferrari:2004ex}.
Supersymmetry has also been claimed to avoid some naturalness problems\,\cite{jain:2005as}
in Lorentz violating theories (but not in the case of noncommutative
models as discussed in\,\cite{ferrari:2005ng}). This discussion
of naturalness in the context of LV in (commutative) supersymmetric
models was developed on more general grounds in\,\cite{grootnibbelink:2004za,bolokhov:2005cj},
with the conclusion that supersymmetry forbids large Lorentz violating
quantum corrections that could become phenomenologically problematic.
As a matter of fact, the supersymmetry algebra is highly constrained,
and as a consequence supersymmetric models themselves are also very
constrained regarding possible terms in the classical Lagrangian and
the quantum corrections, which is the main reason that some problematic
contributions are avoided in many of the contexts mentioned above.
For the same reason, consistent deformations of the supersymmetry
algebra are not trivial do obtain, some noteworthy examples involving
the use of Hopf algebras\,\cite{kosinski:1994vc,kobayashi:2004ep,Irisawa:2006xx,Dimitrijevic:2009mt,Dimitrijevic:2011zg,Palechor:2016rkq}.

Another possibility, presented in\,\cite{Berger:2001rm}, is to deform
supersymmetry by the introduction of a constant background tensor
$k_{\mu\nu}$, by modifying the usual supercharge (in four spacetime
dimensions)
\begin{equation}
Q_{a}=\partial_{a}-i\bar{\theta}^{\dot{a}}\sigma_{a\dot{a}}^{\mu}\partial_{\mu}\thinspace,\label{eq:1}
\end{equation}
by means of the substitution
\begin{equation}
\partial_{\mu}\rightarrow\partial_{\mu}+k_{\mu\nu}\partial^{\nu}\thinspace,
\end{equation}
where, hereafter, greek and latin indices are used to represent, respectively,
spacetime and spinor indices. In this way, it is possible to introduce
LV in supersymmetric models preserving most of the supersymmetric
structure. The superalgebra is deformed to
\begin{equation}
\left\{ Q_{\alpha},\overline{Q}_{\dot{\beta}}\right\} =\sigma_{\alpha\dot{\beta}}^{\mu}2i\left(\partial_{\mu}+k_{\mu\nu}\partial^{\nu}\right)\thinspace,
\end{equation}
and one may still define superfields, and calculate quantum corrections
using supergraphs. For the specific choice
\begin{equation}
k_{\mu\nu}=\alpha k_{\mu}k_{\nu}\thinspace,
\end{equation}
the resulting deformed models are known as \char`\"{}aether-like\char`\"{},
where now the constant vector $k_{\mu}$ defines a preferred direction
in spacetime. Some quantum properties of these models in three and
four spacetime dimensions were studied in\,\cite{Farias:2012ed,Lehum:2013pca},
both for scalar and gauge theories, by using the fact that the essential
structure of supergraph calculations is preserved by the deformation\,\eqref{eq:1}. 

The study of lower-dimensional theories is motivated by the fact that
many models can be constructed presenting interesting properties,
sometimes resembling those of more complicated four-dimensional models,
but in a simpler setting. On the other hand, unique results can also
be obtained, as the presence of a topological mass for the gauge field
in Chern-Simons models in three spacetime dimensions\,\cite{Deser:1981wh}.
The two-dimensional quantum electrodynamics, also known as Schwinger
model\,\cite{Schwinger:1962tp}, exhibits confinement and mass generation
for the photon, properties similar to what it is expected to happen
in QCD, but which can be understood in detail since the Schwinger
model can be solved exactly\,\cite{abdalla2D}. When supersymmetry
is included, the improvement in the ultraviolet behavior can lead
even to the formulation of finite theories with minimal supersymmetry\,\cite{Dudal:2006ip,ferrari:2007mh,ferrari:2007vv}. 

It is therefore interesting to incorporate aether-like Lorentz violation
in lower dimensional supersymmetric models, to see how their properties
are affected by the presence of the preferred direction in spacetime
defined by $k^{\mu}$. A first step in this direction was presented
in\,\cite{Lehum:2015dqr}, where the aether-like Schwinger model
was studied, showing that the background $k^{\mu}$ affects the pole
structure of the propagators, and that the perturbative mass generation
for the photon is also affected by the constant background vector,
since the generated mass is of the form
\begin{equation}
M=e^{2}\Delta/\left(8\pi^{2}\right)\thinspace,
\end{equation}
where $\Delta=\left(1+\alpha k^{\mu}k_{\mu}\right)^{-1}$. It is noteworthy
that the dependence of $M$ in the LV parameter $k^{\mu}$ is such
that the mass does not depend on the direction of $k^{\mu}$, in the
sense that $k^{\mu}k_{\mu}$ is a scalar from the observer point of
view (and a set of independent scalars from the particle point of
view). A similar result was found when looking from LV contributions
to axions physics in\,\cite{Borges:2013eda}.

In this work we study the incorporation of aether-like Lorentz violation
in the $CP^{\left(N-1\right)}$ model. Essentially, a $CP^{\left(N-1\right)}$
model contains a set of $N$ scalars $\varphi_{i}$, on which it is
imposed a constraint of the form $\left|\vec{\varphi}\right|^{2}=\text{const}$.
On the classical level, the model is globally $SU(N)$ invariant,
and a local $U(1)$ gauge symmetry can be incorporated. Several interesting
properties of such models have been reported in the literature, including
a nontrivial phase structure\,\cite{Arefeva:1980ms}, instantons
solutions and confinement~\cite{DAdda:1978vbw,DAdda:1978dle}. The
coupling with fermions preserves the phase structure but the long
range force is obstructed by the fermionic fields\,\cite{Abdalla:1990qf}.
More recently, the $CP^{\left(N-1\right)}$ model, both with and without
supersymmetry, was studied in the context of spacetime noncommutativity,
and questions like renormalizability\,\cite{asano:2003ix,asano:2004vq,ferrari:2006xx},
and the structure of BPS and non-BPS solitons\,\cite{Lee:2000ey,Furuta:2002nv,Furuta:2002ty,Foda:2002nt,Otsu:2003fq,Ghosh:2003ka,Otsu:2004fz}
were addressed. 

This paper is organized as follows. In Section~\ref{sec2}, we present
the aether-superfield formulation of the model and study its phase
structure, which we will se is not modified by the Lorentz violation.
Leading order corrections to the effective action of the auxiliary
and gauge superfields are calculated in Section~\ref{sec3}, and
the corresponding dispersion relations are discussed, showing that
in general the dynamics is consistent if $\alpha$ is assumed to be
small. In Section~\ref{sec4} we evaluate the quadratic part of the
effective action of the scalar superfield at the subleading order.
Finally, in the Section~\ref{conc} we present our final remarks.

\section{\label{sec2}$CP^{(N-1)}$ model in aether-superspace and its phase
structure}

The two-dimensional aether-superspace was presented in\,\cite{Lehum:2015dqr},
revealing straight similarity with the three-dimensional one: the
conventions of three-dimensional superspace can be directly applied
to the two-dimensional one\,\cite{Gomes:2011aa}, therefore we are
adopting the superspace conventions as described in\,\cite{Gates:1983nr}
for the three-dimensional case, and we will be able to develop a discussion
that encompasses both two and three spacetime dimensions. 

The $CP^{(N-1)}$ model defined in the aether-superspace is a model
$SU(N)$ globally and $U(1)$ supergauge invariant, involving an $N$-uple
of complex scalar aether-superfields $\Phi_{a}$, subject to the constraint
$\Phi_{a}^{*}\Phi_{a}=N/g$, where $g$ is a constant. We refer the
reader to\,\cite{Farias:2012ed,Lehum:2015dqr} for more details on
the modification in the superfield component structure induced by
the presence of the aether. The constraint can be implemented by the
use of a Lagrange multiplier $\Sigma$, which is a real scalar aether-superfield.
As a consequence, the Lorentz-violating supersymmetric $CP^{(N-1)}$
model in the aether-superspace can be defined by the action 
\begin{equation}
S=-\int d^{D}xd^{2}\theta\Big{\{}~{\frac{1}{2}}~\overline{\tilde{\nabla}^{\alpha}\Phi_{a}}\tilde{\nabla}_{\alpha}\Phi_{a}+\Sigma\Big(\Phi_{a}\bar{\Phi}_{a}-{\frac{N}{g}}\Big)\Big{\}}\,,\label{c3eq2}
\end{equation}
where $D$ is the dimension of spacetime, which in this work will
be taken as $D=2$ or $D=3$, 
\begin{equation}
\tilde{\nabla}_{\alpha}=\tilde{D}_{\alpha}-i\Gamma_{\alpha}
\end{equation}
is the supercovariant spinorial derivative, $\tilde{D}_{\alpha}$
being defined as 
\begin{equation}
\tilde{D}_{\alpha}=\partial_{\alpha}+i{\theta}^{{\beta}}(\gamma^{m})_{{\beta}\alpha}\tilde{\partial}_{m}\thinspace,
\end{equation}
with 
\begin{equation}
\tilde{\partial}_{m}=\partial_{m}+k_{mn}\partial^{n}\hat{A}\thinspace.
\end{equation}
Lorentz violation is described by the tensor $k_{mn}=\alpha k_{m}k_{n}$,
where $k^{m}$ is a constant vector with $k^{m}k_{m}$ being equal
either to $1$, $-1$ or $0$, and $\alpha$ is small\,\cite{Carroll:2008pk,Gomes:2009ch}.
The gauge aether-superfield $\Gamma_{\alpha}$ has no kinetic term
in the classical Lagrangian so it has no propagating degrees of freedom
at the classical level, in the same way as the Lagrange multiplier
$\Sigma$, but both acquire nontrivial dynamics at the quantum level.

The usual $CP^{(N-1)}$ model presents a rich phase structure~\cite{Arefeva:1980ms}:
in one phase the symmetry $SU(N)$ is broken down to $SU(N-1)$, whereas
in the other the model remains $SU(N)$ symmetric while mass generation
is observed for the fundamental bosonic fields. We will verify the
possibility of symmetry breaking in our case by assuming that $\Sigma$
and the $a=N$ component of $\Phi_{a}$ acquire non-vanishing vacuum
expectation values (VEVs), 
\[
\langle\Sigma\rangle=m\text{ and }\langle\Phi_{N}\rangle=\sqrt{N}v\thinspace.
\]
By redefining the aether-superfields in term of new ones with vanishing
VEV, $\Phi_{a}\longrightarrow\Phi_{a}$ (for $a=1,\ldots,N-1$), $\Phi_{N}\longrightarrow\Phi_{N}+{v\sqrt{N}}$,
$\bar{\Phi}_{N}\longrightarrow\bar{\Phi}_{N}+{\bar{v}\sqrt{N}}$,
and $\Sigma\longrightarrow\Sigma+m$, the action in Eq.\,(\ref{c3eq2})
can be cast as 
\begin{eqnarray}
S & = & \int{d^{D}xd^{2}\theta}\Big\{\bar{\Phi}_{a}(\tilde{D}^{2}-m)\Phi_{a}-\Sigma(\bar{\Phi}_{a}\Phi_{a}-\frac{N}{g}+N~v\bar{v})-\frac{1}{2}\bar{\Phi}_{a}\Phi_{a}\Gamma^{\alpha}\Gamma_{\alpha}\nonumber \\
 & + & \frac{i}{2}\left(\tilde{D}^{\alpha}\bar{\Phi}_{a}\Phi_{a}\Gamma_{\alpha}-\bar{\Phi}_{a}D^{\alpha}\Phi_{a}\Gamma_{\alpha}+v\sqrt{N}\tilde{D}^{\alpha}\bar{\Phi}_{N}\Gamma_{\alpha}-\bar{v}\sqrt{N}\Gamma^{\alpha}\tilde{D}_{\alpha}\Phi_{N}\right)\nonumber \\
 & - & \frac{\sqrt{N}}{2}\bar{v}\Phi_{N}\Gamma^{\alpha}\Gamma_{\alpha}+\frac{\sqrt{N}}{2}v\bar{\Phi}_{N}\Gamma^{\alpha}\Gamma_{\alpha}+\frac{N}{2}\bar{v}v\Gamma^{\alpha}\Gamma_{\alpha}-m\sqrt{N}(v\bar{\Phi}_{N}+\bar{v}\Phi_{N})\nonumber \\
 & - & \Sigma\Phi_{N}\bar{v}\sqrt{N}+\Sigma\bar{\Phi}_{N}v\sqrt{N}\Big\}.
\end{eqnarray}
The propagator for the first $(N-1)$ components of $\Phi_{a}$ is
readily seen to be given by 
\begin{equation}
\langle T\,\Phi_{a}(\tilde{p},\theta_{1})\bar{\Phi}_{b}(-\tilde{p},\theta_{2})\rangle=-i\delta_{ab}\left(\frac{\tilde{D}^{2}+m}{\tilde{p}^{2}+m^{2}}\right)\delta^{2}(\theta_{1}-\theta_{2}).\label{c3eq9a}
\end{equation}

\begin{figure}[h]
\includegraphics{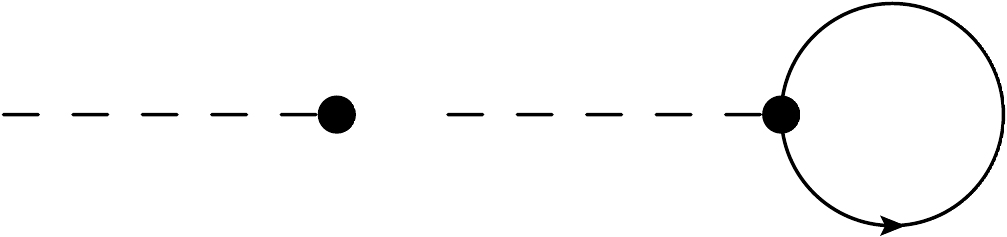} \caption{Gap equation for the $\Sigma$ aether-superfield, where solid lines
represent $\Phi_{a}$ propagators and dashed lines represent the $\Sigma$
propagators.}
\label{Fig1} 
\end{figure}

The redefined aether-superfields must have vanishing vacuum expectation
values, $\langle\Sigma\rangle=0$ and $\langle\Phi_{N}\rangle=0$,
implying in the equations 
\begin{align}
 & i\int\frac{d^{D}q}{(2\pi)^{D}}\frac{1}{\tilde{q}^{2}+m^{2}}+\frac{1}{g}+\delta_{g}-v\bar{v}=0\thinspace,\label{cond01}\\[0.5cm]
 & m\bar{v}=mv=0\thinspace,\label{cond02}
\end{align}
which are represented in Fig.\,\ref{Fig1}. From Eq.(\ref{cond02}),
we see that for $v=0$ and $m\neq0$ a non vanishing mass is generated
for the fundamental aether-superfields $\Phi_{a}$, in the gauge symmetric
phase. In such a case, Eq. (\ref{cond01}) reads 
\begin{equation}
\delta_{g}=-\frac{1}{g}-i\int\frac{d^{D}q}{(2\pi)^{D}}\frac{1}{\tilde{q}^{2}+m^{2}}\thinspace,\label{cond03}
\end{equation}
corresponding to the renormalization of the coupling constant $g$.
Notice that in two dimensions the above integral is logarithmic divergent,
while in three dimensions is finite because the integral (linear UV
divergent) is completely regularized by dimensional reduction. For
the other phase, we have $v\neq0$ and $m=0$, implying that the global
symmetry is reduced to $SU\left(N-1\right)$, while the scalar aether-superfields
remain massless. We therefore conclude that the presence of the aether-like
Lorentz violation does not change the phase structure of the $CP^{(N-1)}$
model.

\section{\label{sec3}Effective action at leading order}

We shall be interested in the generation of mass in the model, so
we will work in the symmetric phase from now on. Moreover, the renormalization
of the model is not affected by the choice of phase because both have
the same ultraviolet behavior~\cite{coleman1988aspects}. Therefore,
in the symmetric phase ($v=0$ and $m\neq0$) the action reads 
\begin{eqnarray}
S & = & \int{d^{D}xd^{2}\theta}\Big\{\bar{\Phi}_{a}(\tilde{D}^{2}-m)\Phi_{a}-\Sigma\left(\Phi_{a}\bar{\Phi}_{a}-\frac{N}{g}\right)\nonumber \\
 &  & -\frac{1}{2}\bar{\Phi}_{a}\Phi_{a}\Gamma^{\alpha}\Gamma_{\alpha}+\frac{i}{2}\left[\tilde{D}^{\alpha}\bar{\Phi}_{a}\Phi_{a}\Gamma_{\alpha}-\bar{\Phi}_{a}\tilde{D}^{\alpha}\Phi_{a}\Gamma_{\alpha}\right]\Big\},
\end{eqnarray}
where $a=1,\cdots,N$; in this case, the propagator of the $\Phi_{N}$
aether-superfield is also given by Eq. (\ref{c3eq9a}).

\subsection{\label{tpsigma}Effective action of the $\Sigma$ aether-superfield}

At the classical level, the $\Sigma$ aether-superfield is a constraint
without dynamics, however it acquires a nonlocal kinetic term becoming
a propagating dynamical superfield by means of quantum corrections.
In the leading order of the $1/N$ expansion, the radiative correction
to its quadratic effective action, depicted in Fig.\,\ref{Fig2},
is given by 
\begin{equation}
\Gamma_{\Sigma}^{(2)}=\frac{1}{2}\int\frac{d^{D}p}{(2\pi)^{D}}d^{2}\theta~\Sigma(-\tilde{p},\theta)\left[N~f(\tilde{p})(\tilde{D}^{2}+2m)\right]\Sigma(\tilde{p},\theta),\label{c3eq20}
\end{equation}
where 
\begin{equation}
f(\tilde{p})=\int\frac{d^{D}q}{(2\pi)^{D}}\frac{1}{[(\tilde{q}+\tilde{p})^{2}+m^{2}](\tilde{q}^{2}+m^{2})}\thinspace,\label{c3eq21}
\end{equation}
and the computer package presented in\,\cite{ferrarisusymath} were
used for the manipulations of covariant superderivatives. In order
to evaluate the integral, we use the Feynman trick to combine the
denominators, and calculate the integral in $D$ dimensions obtaining
\begin{eqnarray}
f(\tilde{p}) & = & \int\frac{d^{D}q}{(2\pi)^{D}}\frac{1}{[(\tilde{q}+\tilde{p})^{2}+m^{2}](\tilde{q}^{2}+m^{2})}=\Delta\int_{0}^{1}{dz}\int\frac{d^{D}q}{(2\pi)^{D}}\frac{1}{(\tilde{q}^{2}+M)^{2}}\nonumber \\
 & = & \Delta\int_{0}^{1}{dz}~\frac{i}{(4\pi)^{D/2}}\frac{\Gamma\left(2-D/2\right)}{\Gamma(2)}\left(\frac{1}{M}\right)^{2-D/2},
\end{eqnarray}
where $M=m^{2}+\tilde{p}^{2}z(1-z)$. The factor
\begin{equation}
\Delta=\det\left(\frac{\partial q^{m}}{\partial\tilde{q}^{n}}\right)=\det{}^{-1}(\delta_{n}^{m}+k_{n}^{m})\thinspace,
\end{equation}
is the Jacobian of the change of variables $q\rightarrow\tilde{q}$
used to bring the integral to its final form. 

In particular, for $D=2$, $f(\tilde{p})$ reads 
\begin{eqnarray}
f(\tilde{p}) & = & \Delta\int_{0}^{1}{dz}~\left\{ \frac{1}{4\pi[m^{2}+\tilde{p}^{2}z(1-z)]}+\mathcal{O}(D-2)\right\} \nonumber \\
 & = & -\frac{\Delta\text{arctan}\left(|\tilde{p}|/{\sqrt{-4m^{2}-\tilde{p}^{2}}}\right)}{\pi|\tilde{p}|\sqrt{-4m^{2}-\tilde{p}^{2}}}=\left\{ \begin{array}{ll}
\Delta\ln{(\tilde{p}^{2}/m^{2})}/(2\pi\tilde{p}^{2})\hspace{0.5cm}{\rm for}~~~\tilde{p}\rightarrow~\infty\\
\Delta/(4\pi m^{2})\hspace{2.3cm}{\rm for}~~~\tilde{p}\rightarrow~0
\end{array}\right.,\label{eq:ftilde2d}
\end{eqnarray}

\noindent while in $D=3$, 
\begin{eqnarray}
f(\tilde{p}) & = & \Delta\int_{0}^{1}{dz}~\left\{ \frac{1}{8\pi\sqrt{m^{2}+\tilde{p}^{2}z(1-z)}}+\mathcal{O}(D-3)\right\} \nonumber \\
 & = & \frac{\Delta}{8\pi|\tilde{p}|}\ln\left(\frac{2|m|-i|\tilde{p}|}{2|m|+i|\tilde{p}|}\right)=\left\{ \begin{array}{ll}
\Delta/(8|\tilde{p}|)\hspace{1cm}{\rm for}~~~\tilde{p}\rightarrow~\infty\\
\Delta/(8\pi|m|)\hspace{0.6cm}{\rm for}~~~\tilde{p}\rightarrow~0
\end{array}\right..\label{eq:ftilde3d}
\end{eqnarray}
From Eq.\,(\ref{c3eq20}), we obtain the radiative induced $\Sigma$
propagator, 
\begin{eqnarray}
\langle T\,\Sigma(\tilde{p},\theta_{1})\Sigma(-\tilde{p},\theta_{2})\rangle=\frac{i}{N}\frac{(\tilde{D}^{2}-2m)}{f(\tilde{p})(\tilde{p}^{2}+4m^{2})}\delta(\theta_{1}-\theta_{2}).\label{c3eq22}
\end{eqnarray}

\noindent This propagator has a regular infrared behavior ($\tilde{p}^{2}\rightarrow0$)
in both dimensions, while it decreases as $1/|\tilde{p}|$ for $D=3$
and $1/\ln|\tilde{p}|$ for $D=2$ in the ultraviolet limit ($p^{2}\rightarrow\infty$).
We notice that the nonlocal $f(\tilde{p})$ factor in the denominator
does not introduce additional poles, so the dispersion relation for
the aether-superfield $\Sigma$ is given by 
\begin{equation}
\tilde{p}^{2}+4m^{2}=p^{2}+2k_{mn}p^{m}p^{n}+k^{mn}k_{ml}p_{n}p^{l}+4m^{2}=0\thinspace,
\end{equation}
or
\begin{equation}
p^{2}+2\alpha\left[1+\alpha k^{2}\right]\left(k^{\mu}p_{\mu}\right)^{2}+4m^{2}=0\thinspace,\label{eq:DR}
\end{equation}
in terms of the aether LV vector $k^{\mu}$. 

It is interesting to analyze the consequences of Eq.\,\eqref{eq:DR}
separately for $k^{\mu}k_{\mu}$ being $\pm1$ or zero, as it was
done in\,\cite{Farias:2012ed}. Starting with spacelike $k^{\mu}$,
i.e., $k^{\mu}k_{\mu}=+1$, we choose coordinates such that $k^{\mu}=\left(0,\hat{k}\right)$,
$\hat{k}$ being a unitary space vector. With this, Eq.\,\eqref{eq:DR}
reduces to
\begin{equation}
E^{2}=\vec{p}^{2}+4m^{2}+\alpha\left(2+\alpha\right)\left(\hat{k}\cdot\vec{p}\right)^{2}\thinspace,
\end{equation}
and, by taking $\vec{p}=\vec{0}$, the rest mass of the particle if
found to be $2m$, independent of $\alpha$ and $\hat{k}$ and therefore
of the LV background. Generally, however, the particle energy depends
on the orientation of its momentum with respect to the LV vector $\hat{k}$.
We may also notice that if $\alpha=-1$ and $\vec{p}$ collinear to
$\hat{k}$, w obtain $E^{2}=4m^{2}$, independent of $\vec{p}$, therefore
representing a degenerate, inconsistent dynamics. Otherwise, for any
small value of $\alpha$, the dynamics is consistent. Now considering
timelike $k^{\mu}$, i.e., $k^{\mu}k_{\mu}=-1$, we choose coordinates
such that $k^{\mu}=\left(1,\vec{0}\right)$, thus obtaining
\begin{equation}
E^{2}=\frac{\vec{p}^{2}+4m^{2}}{1-\alpha^{2}}\thinspace,
\end{equation}
corresponding to a rest mass given by $2m/\sqrt{1-\alpha^{2}}$. In
this case, we see an explicit dependence of the dynamics (which is
consistent in principle as far as $\left|\alpha\right|<1$) on the
LV properties, as given by the value of $\alpha$. Finally, for the
lightlike case, we have a more complicated dispersion relation,
\begin{equation}
E^{2}\left[1-2\alpha\left(k^{0}\right)^{2}\right]+4\alpha k^{0}\left(\vec{k}\cdot\vec{p}\right)E-2\alpha\left(\vec{k}\cdot\vec{p}\right)^{2}-\vec{p}^{2}-4m^{2}=0\thinspace.
\end{equation}
To simplify matters, the reference frame is chosen so that $\vec{k}=\left(k^{0},k^{0},0,0\right)$.
We consider two particular cases: first, $\vec{p}$ parallel to $\vec{k}$,
leading to
\begin{equation}
E=\left(1-2\alpha\right)^{-1}\left[-2\alpha\left|\vec{p}\right|\pm\sqrt{\vec{p}^{2}+4m^{2}\left(1-2\alpha\right)}\right]\thinspace,
\end{equation}
and also $\vec{p}$ perpendicular to $\vec{k}$, in which case we
obtain
\begin{equation}
E^{2}=\frac{\vec{p}^{2}+4m^{2}}{1-2\alpha}\thinspace.
\end{equation}
We see that the result for $\vec{p}\perp\vec{k}$ is very similar
to the timelike case studied before. For $\vec{p}\parallel\vec{k}$,
given a sufficiently small value of $\left|\alpha\right|$, we have
one positive and one negative energy states, which are to be reinterpreted,
in the quantum theory, as the energies for particles and antiparticles.
In this case, it is noteworthy that the energies of particles and
their corresponding antiparticles will be slightly different. Finally,
both for parallel and perpendicular cases, the rest mass is given
by $2m/\sqrt{1-2\alpha}$.

\subsection{Effective action of the gauge aether-superfield}

\begin{figure}[h]
\includegraphics{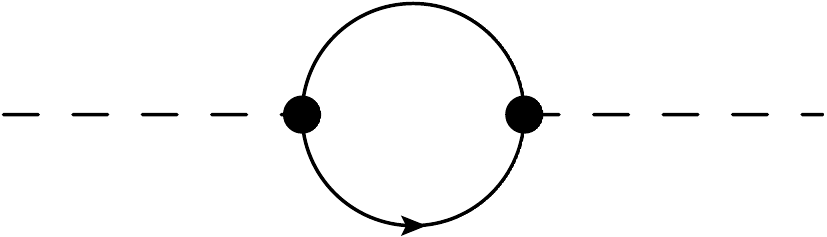} \caption{Contribution to the effective propagator of the $\Sigma$ aether-superfield
at leading order of $1/N$.}
\label{Fig2} 
\end{figure}

\begin{figure}[h]
\includegraphics{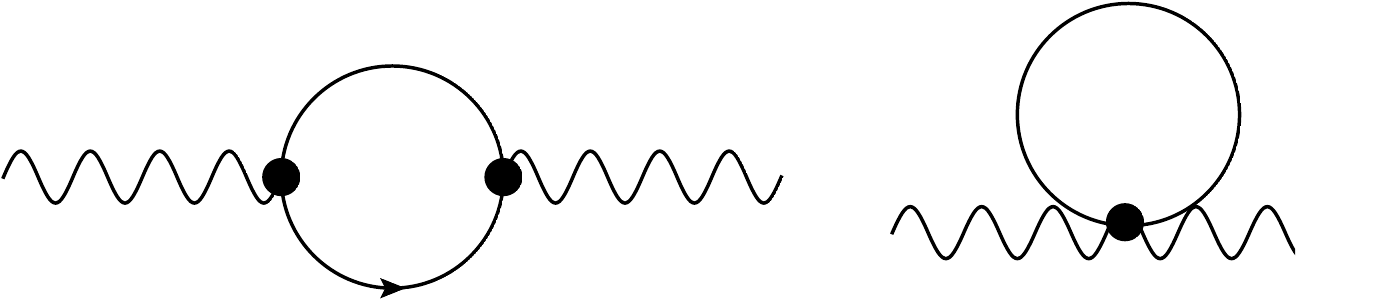} \caption{Contribution to the gauge aether-superfield effective propagator at
leading order of $1/N$. Wavy lines represent the $\Gamma_{\alpha}$
superpropagator.}
\label{Fig3} 
\end{figure}

The gauge aether-superfield $\Gamma_{\alpha}$ is also non dynamical
at classical level, but similarly to $\Sigma$ it will have a nontrivial
kinetic term generated by quantum corrections. The leading $1/N$
radiative corrections to the quadratic part of the $\Gamma_{\alpha}$
effective action are represented in Fig.\,\ref{Fig3}. The first
contribution, Fig.\ref{Fig3}a, is given by 
\begin{equation}
\Gamma_{\ref{Fig2}a}^{(2)}=-\frac{N}{2}\int\frac{d^{D}p}{(2\pi)^{D}}d^{2}\theta~\Gamma_{\beta}(-\tilde{p},\theta)\int\frac{d^{D}\tilde{q}\Delta}{(2\pi)^{D}}\frac{C^{\alpha\beta}}{\tilde{q}^{2}+m^{2}}\Gamma_{\alpha}(\tilde{p},\theta),\label{c3eq23c}
\end{equation}

\noindent while for the diagram \ref{Fig3}b, 
\begin{eqnarray}
\Gamma_{\ref{Fig2}b}^{(2)} & = & \frac{N}{2}\int\frac{d^{D}p}{(2\pi)^{D}}d^{2}\theta~\Gamma_{\beta}(-\tilde{p},\theta)\int\frac{d^{D}\tilde{q}\Delta}{(2\pi)^{D}}\frac{1}{[(\tilde{q}+\tilde{p})^{2}+m^{2}](\tilde{q}^{2}+m^{2})}\nonumber \\
 & \times & \left[(\tilde{q}^{2}+m^{2})C^{\alpha\beta}+(\tilde{q}^{\alpha\beta}+mC^{\alpha\beta})\tilde{D}^{2}+\frac{1}{2}(\tilde{q}^{\gamma\beta}+mC^{\gamma\beta})\tilde{D}_{\gamma}\tilde{D}^{\alpha}\right]\Gamma_{\alpha}(\tilde{p},\theta).
\end{eqnarray}

\noindent Summing up the two contributions above we have 
\begin{eqnarray}
\Gamma_{\ref{Fig2}}^{(2)} & = & \frac{N}{2}\int\frac{d^{D}p}{(2\pi)^{D}}d^{2}\theta~\Gamma_{\beta}(-\tilde{p},\theta)\int\frac{d^{D}\tilde{q}\Delta}{(2\pi)^{D}}\frac{1}{[(\tilde{q}+\tilde{p})^{2}+m^{2}](\tilde{q}^{2}+m^{2})}\nonumber \\
 & \times & \left[(\tilde{q}^{\alpha\beta}+mC^{\alpha\beta})\tilde{D}^{2}+\frac{1}{2}(\tilde{q}^{\gamma\beta}+mC^{\gamma\beta})\tilde{D}_{\gamma}\tilde{D}^{\alpha}\right]\Gamma_{\alpha}(\tilde{p},\theta).
\end{eqnarray}
Using the relation $\tilde{D}_{\mu}\tilde{D}_{\nu}=i\tilde{\partial}_{\mu\nu}+C_{\nu\mu}\tilde{D}^{2}$,
$\Gamma_{\ref{Fig2}}^{(2)}$ can be rewritten as 
\begin{equation}
\Gamma_{\ref{Fig2}}^{(2)}=\frac{N}{2}\int\frac{d^{D}p}{(2\pi)^{D}}d^{2}\theta~\Gamma_{\beta}(-\tilde{p},\theta)\int\frac{d^{D}\tilde{q}\Delta}{(2\pi)^{D}}\frac{\tilde{q}^{\beta\gamma}-mC^{\beta\gamma}}{[(\tilde{q}+\tilde{p})^{2}+m^{2}](\tilde{q}^{2}+m^{2})}W_{\gamma}(\tilde{p},\theta)\thinspace,\label{c3eq24b}
\end{equation}
where $W^{\alpha}=\frac{1}{2}D^{\beta}D^{\alpha}\Gamma_{\beta}$ is
the Maxwell aether-superfield strength.

Using the identity 
\begin{equation}
\int\frac{d^{D}q}{(2\pi)^{D}}\frac{q^{\alpha\beta}}{[(q+p)^{2}+m^{2}](q^{2}+m^{2})}=-\frac{p^{\alpha\beta}}{2}\int\frac{d^{D}q}{(2\pi)^{D}}\frac{1}{[(q+p)^{2}+m^{2}](q^{2}+m^{2})}\thinspace,\label{c3eq25}
\end{equation}
Eq.\,(\ref{c3eq24b}) can be cast as 
\begin{eqnarray}
\Gamma_{\ref{Fig2}}^{(2)} & = & -\frac{N}{2}\int\frac{d^{D}p}{(2\pi)^{D}}d^{2}\theta~f(\tilde{p})\left[W^{\alpha}(-\tilde{p},\theta)W_{\alpha}(\tilde{p},\theta)+2m\Gamma^{\alpha}(-\tilde{p},\theta)W_{\alpha}(\tilde{p},\theta)\right].\label{c3eq26}
\end{eqnarray}
In $D=3$, the previous effective action describes the dynamics of
a (non-local) Maxwell-Chern-Simons aether-superfield. It is well-known
that the presence of Chern-Simons (CS) term $\Gamma^{\alpha}W_{\alpha}$
generates a topological massive pole for the $\Gamma_{\alpha}$ two-point
superpropagator. In $D=2$, the effective action we obtained represents
the dynamics of a massive gauge invariant aether-superfield, as discussed
in the ordinary two dimensional superspace in\,\cite{Gates:1977hb,Bengtsson:1983in}.

The propagator of the $\Gamma_{\alpha}$ aether-superfield at leading
order can be obtained after a gauge fixing. For convenience, we use
a covariant nonlocal gauge fixing, together with the corresponding
Faddeev-Popov terms, given by 
\begin{eqnarray}
S_{GF} & = & \frac{N}{4}\int\frac{d^{D}p}{(2\pi)^{D}}d^{2}\theta f(\tilde{p})\left\{ \frac{1}{2\xi}\tilde{D}^{\beta}\Gamma_{\beta}\tilde{D}^{2}\tilde{D}^{\alpha}\Gamma_{\alpha}-c^{\prime}\tilde{D}^{2}c\right\} .\label{c3eq31}
\end{eqnarray}
Notice that the ghosts aether-superfields decouple from the $\Gamma_{\alpha}$
because we are working in an Abelian gauge theory. With this gauge
choice, we obtain the following gauge aether-superfield propagator
\begin{equation}
\langle T~\Gamma^{\alpha}(\tilde{p},\theta_{1})\Gamma^{\beta}(-\tilde{p},\theta_{2})\rangle=\frac{i}{Nf(\tilde{p})}\left[\frac{(\tilde{D}^{2}-2m)\tilde{D}^{\beta}\tilde{D}^{\alpha}}{\tilde{p}^{2}(\tilde{p}^{2}+4m^{2})}+\xi\frac{\tilde{D}^{2}\tilde{D}^{\alpha}\tilde{D}^{\beta}}{(\tilde{p}^{2})^{2}}\right]\delta(\theta_{1}-\theta_{2}).\label{c3eq33}
\end{equation}
The pole at $p^{2}=0$ is not physical, being dependent on the gauge
parameter $\xi$. To verify this, the standard procedure is to project
the superpropagator given in Eq.\,\eqref{c3eq33} for the physical
component fields. For example, for $2+1$ spacetime dimensions, we
use the decomposition
\begin{equation}
\Gamma_{\alpha}(\tilde{p},\theta_{1})=\chi_{\alpha}-\theta_{\alpha}B+i\theta^{\beta}V_{\alpha\beta}-\theta^{2}\left(\lambda_{\alpha}+i\partial_{\alpha\beta}\chi^{\beta}\right)\thinspace,
\end{equation}
together with the selection of the Wess-Zumino gauge, $B=\chi=0$,
to notice that
\begin{equation}
\langle T~\Gamma^{\alpha}(\tilde{p},\theta_{1})\Gamma^{\beta}(-\tilde{p},\theta_{2})\rangle=-\theta_{1}^{\mu}\theta_{2}^{\nu}\langle T~V_{\mu\alpha}(\tilde{p})V_{\nu\beta}(\tilde{p})\rangle+\cdots\thinspace,
\end{equation}
where $\langle T~V_{\mu\alpha}(\tilde{p})V_{\nu\beta}(\tilde{p})\rangle$
is the desired propagator for the gauge field $V_{\mu\alpha}$. To
obtain its explicit form, one starts by calculating explicitly $D^{2}\delta(\theta_{1}-\theta_{2})$
and $\left(D^{2}\right)^{2}\delta(\theta_{1}-\theta_{2})$, 
\begin{align}
D^{2}\delta(\theta_{1}-\theta_{2})= & -1+\theta_{1}^{\mu}\theta_{2}^{\nu}\tilde{p}_{\mu\nu}+\frac{1}{4}\theta_{1}^{\mu}\theta_{1}^{\nu}\theta_{2}^{\alpha}\theta_{2\alpha}\tilde{p}_{\mu}^{\hphantom{\mu}\beta}\tilde{p}_{\beta\nu}\thinspace,\\
\left(D^{2}\right)^{2}\delta(\theta_{1}-\theta_{2})= & -\frac{1}{2}\theta_{1}^{\mu}\theta_{1}^{\nu}\tilde{p}_{\mu}^{\hphantom{\mu}\beta}\tilde{p}_{\beta\nu}+\theta_{1}^{\mu}\theta_{2}^{\nu}\tilde{p}_{\mu}^{\hphantom{\mu}\beta}\tilde{p}_{\beta\nu}-\frac{1}{4}\theta_{2}^{\mu}\theta_{2\mu}\tilde{p}^{\alpha\beta}\tilde{p}_{\alpha\beta}\nonumber \\
 & +\frac{1}{2}\theta_{1}^{\mu}\theta_{1}^{\nu}\theta_{2}^{\alpha}\theta_{2\alpha}\tilde{p}_{\mu}^{\hphantom{\mu}\beta}\tilde{p}_{\beta}^{\hphantom{\beta}\gamma}\tilde{p}_{\gamma\nu}\thinspace.
\end{align}
The identity $\tilde{D}^{\beta}\tilde{D}^{\alpha}=\tilde{p}^{\alpha\beta}+C^{\alpha\beta}D^{2}$
can be used to rewrite Eq.\,\eqref{c3eq33} in terms of $D^{2}$
and $D^{4}$ only, and using the results of the last equation to isolate
the required part, i.e.,
\begin{equation}
\langle T~V_{\mu\alpha}(\tilde{p})V_{\nu\beta}(\tilde{p})\rangle=\left.\langle T~\Gamma_{\alpha}(\tilde{p},\theta_{1})\Gamma_{\beta}(-\tilde{p},\theta_{2})\rangle\right|_{\theta^{\mu}\theta^{\nu}}\thinspace,
\end{equation}
one obtains the final result as
\begin{equation}
\langle T~V_{\mu\alpha}(\tilde{p})V_{\nu\beta}(\tilde{p})\rangle=\frac{F_{\mu\nu\alpha\beta}\left(\tilde{p}\right)}{\tilde{p}^{2}\left(\tilde{p}^{2}+4m^{2}\right)}+\xi\frac{G_{\mu\nu\alpha\beta}\left(\tilde{p}\right)}{\left(\tilde{p}^{2}\right)^{2}}\thinspace,
\end{equation}
where
\begin{align}
F_{\mu\nu\alpha\beta}\left(\tilde{p}\right) & =-\frac{iN}{f\left(\tilde{p}\right)}\left[\tilde{p}_{\alpha\beta}\tilde{p}_{\rho\sigma}+C_{\alpha\beta}\tilde{p}_{\mu}^{\hphantom{\mu}\rho}\tilde{p}_{\rho\nu}-2m\left(C_{\alpha\beta}\tilde{p}_{\rho\sigma}+C_{\rho\sigma}\tilde{p}_{\alpha\beta}\right)\right]\thinspace,\\
G_{\mu\nu\alpha\beta}\left(\tilde{p}\right) & =-\frac{iN}{f\left(\tilde{p}\right)}\left(\tilde{p}_{\alpha\beta}\tilde{p}_{\rho\sigma}-C_{\alpha\beta}\tilde{p}_{\mu}^{\hphantom{\mu}\rho}\tilde{p}_{\rho\nu}\right)\thinspace.
\end{align}
To calculate the residues at the poles $\tilde{p}^{2}=-4m^{2}$ and
$\tilde{p}^{2}=0$, it is convenient to write
\begin{equation}
\langle T~V_{\mu\alpha}(\tilde{p})V_{\nu\beta}(\tilde{p})\rangle=\frac{\left(\tilde{p}^{2}\right)F_{\mu\nu\alpha\beta}\left(\tilde{p}\right)+\xi\left(\tilde{p}^{2}+4m^{2}\right)G_{\mu\nu\alpha\beta}\left(\tilde{p}\right)}{\left(\tilde{p}^{2}\right)^{2}\left(\tilde{p}^{2}+4m^{2}\right)}\thinspace,
\end{equation}
and it becomes evident that the residue at the $\tilde{p}^{2}=0$
pole depends on the gauge parameter $\xi$, and therefore it is unphysical.
The only physical pole is at $\tilde{p}^{2}=-4m^{2}$, corresponding
to the known mass generation for the gauge field in this model. One
notices also that the specific form of the $F_{\mu\nu\alpha\beta}$
and $G_{\mu\nu\alpha\beta}$ functions is not used in this argument,
so actually the same conclusion could be reached directly from the
form of the superfield propagator. The physical pole has the same
dispersion relation as the $\Sigma$ superfield, as discussed before.
In particular, for spacelike LV parameter $k^{\mu}$, we have shown
that the generated mass is actually independent on the LV background.
This is a difference from the dynamical generation of mass observed
in the two-dimensional quantum electrodynamics in the aether-superspace
($SQED_{2}$), where an explicit dependence on the LV background is
always observed for the generated mass of the gauge fields~\cite{Lehum:2015dqr}.

\section{\label{sec4}Effective action at subleading order}

\begin{figure}[h]
\includegraphics{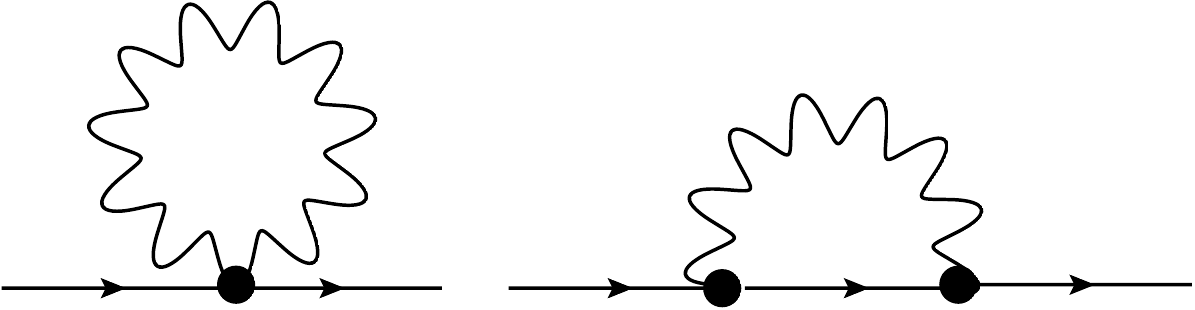} \caption{Feynman diagrams contributing to the $1/N$ subleading corrections
to the quadratic effective action of the $\Phi$ aether-superfield.}
\label{Fig4} 
\end{figure}

We are also able to evaluate the quadratic part of the effective action
of the matter aether-superfield $\Phi$ at sub-leading order, whose
contributions arise from the diagrams depicted in Fig.\,\ref{Fig4}.
The corresponding amplitude is given by 
\begin{align}
\Gamma_{\ref{Fig4}}^{(2)} & =\frac{1}{N}\int\frac{d^{D}p}{(2\pi)^{D}}d^{2}\theta\,\bar{\Phi}(-\tilde{p},\theta)\left(D^{2}-m\right)\Phi(p,\theta)\int\frac{d^{D}\tilde{q}\Delta}{(2\pi)^{D}}\frac{1}{f(\tilde{q})}\nonumber \\
 & \times\left\{ \frac{\tilde{q}^{2}+2\tilde{q}.\tilde{p}}{\tilde{q}^{2}[(\tilde{q}+\tilde{p})^{2}+m^{2}]}\left[\frac{1}{\tilde{q}^{2}+4m^{2}}+\frac{\xi}{\tilde{q}^{2}}\right]-\frac{2m(\tilde{D}^{2}-m)}{\tilde{q}^{2}(\tilde{q}^{2}+4m^{2})[(\tilde{q}+\tilde{p})^{2}+m^{2}]}\right\} \,,
\end{align}
which after some algebraic manipulations, and choosing the gauge $\xi=0$,
leads to
\begin{align}
\Gamma_{\ref{Fig4}}^{(2)} & =\frac{1}{N}\int\frac{d^{D}p}{(2\pi)^{D}}d^{2}\theta\,\bar{\Phi}(-\tilde{p},\theta)\left(D^{2}-m\right)\Phi(p,\theta)\int\frac{d^{D}\tilde{q}\Delta}{(2\pi)^{D}}\frac{1}{f(\tilde{q})}\nonumber \\
 & \times\left\{ \frac{1}{(\tilde{q}^{2}+4m^{2})[(\tilde{q}+\tilde{p})^{2}+m^{2}]}+\frac{(\tilde{D}^{2}-m)^{2}}{\tilde{q}^{2}(\tilde{q}^{2}+4m^{2})[(\tilde{q}+\tilde{p})^{2}+m^{2}]}\right\} \,,
\end{align}
where we can see that, despite being non-local, the quantum correction
to the two-point vertex function of the $\Phi$ aether-superfield
exhibits the familiar $(D^{2}-m)$ factor. The second term in the
brackets of the previous equation turns out to be an UV finite high
order derivative monomial generated by radiative corrections. The
first term in the integral is log-divergent in both dimensions, because
of the asymptotic behavior of $f(\tilde{q})$, Eqs.(\ref{eq:ftilde2d})
and (\ref{eq:ftilde3d}).

\section{\label{conc}Final remarks}

In this paper we have studied the dynamical generation of mass in
the two and three-dimensional Lorentz-violating supersymmetric $CP^{(N-1)}$
model defined in the aether-superspace. Even though the phase structure
of the model is not affected by the Lorentz violation, we showed that
in the $CP^{(N-1)}$ model the dynamically generated mass has an explicit
dependence on the aether properties, except for spacelike LV vector.
The aether properties dependence on the physical dynamically generated
mass has also been noticed in the quantum electrodynamics in the two-dimensional
aether-superspace~\cite{Lehum:2015dqr}. We have also calculated
the leading quantum corrections, in the large $N$ approximation,
to the two-point vertex functions of the scalar, gauge and auxiliary
aether-superfields, showing that the later two acquire dynamics at
the quantum level, and that the generated propagators are sensitive
to the LV. The dispersion relation governing their dynamics was studied,
and shown to be consistent for small $\alpha$. Finally, subleading
corrections to the scalar propagators were also obtained, also exhibiting
a dependence on the aether parameter. 

\textbf{\medskip{}
}

\textbf{Acknowledgments.} This work was supported by Conselho Nacional
de Desenvolvimento Científico e Tecnológico (CNPq) and Fundação de
Amparo a Pesquisa do Estado de São Paulo (FAPESP), via the following
grants: FAPESP 2017/13767-9 (AFF), CNPq 307723/2016-0 and 402096/2016-9
(ACL). \textbf{\medskip{}
}

\begin{thebibliography}{10}

\bibitem{colladay:1998fq}
D.~Colladay and V.~A. Kostelecky.
\newblock {Lorentz-violating extension of the standard model}.
\newblock {\em Phys. Rev.}, D58:116002, 1998, hep-ph/9809521.

\bibitem{Kostelecky:2008ts}
V.~Alan Kostelecky and Neil Russell.
\newblock {Data Tables for Lorentz and CPT Violation}.
\newblock {\em Rev. Mod. Phys.}, 83:11, 2011, 0801.0287.

\bibitem{Terashima:2000xq}
Seiji Terashima.
\newblock {A Note on superfields and noncommutative geometry}.
\newblock {\em Phys.Lett.}, B482:276--282, 2000, hep-th/0002119.

\bibitem{seiberg:1999vs}
N.~Seiberg and E.~Witten.
\newblock {String theory and noncommutative geometry}.
\newblock {\em JHEP}, 09:032, 1999, hep-th/9908142.

\bibitem{girotti:2000gc}
H.~O. Girotti, M.~Gomes, Victor~O. Rivelles, and A.~J. da~Silva.
\newblock {A consistent noncommutative field theory: The Wess-Zumino model}.
\newblock {\em Nucl. Phys.}, B587:299--310, 2000, hep-th/0005272.

\bibitem{ferrari:2003vs}
A.~F. Ferrari, H.~O. Girotti, M.~Gomes, A.~Yu. Petrov, A.~A. Ribeiro, V.~O.
  Rivelles, and A.~J. da~Silva.
\newblock {Superfield covariant analysis of the divergence structure of
  noncommutative supersymmetric {QED(4)}}.
\newblock {\em Phys. Rev.}, D69:025008, 2004, hep-th/0309154.

\bibitem{ferrari:2004ex}
A.~F. Ferrari, H.~O. Girotti, M.~Gomes, A.~Yu. Petrov, A.~A. Ribeiro, V.~O.
  Rivelles, and A.~J. da~Silva.
\newblock {Towards a consistent noncommutative supersymmetric Yang- Mills
  theory: Superfield covariant analysis}.
\newblock {\em Phys. Rev.}, D70:085012, 2004, hep-th/0407040.

\bibitem{jain:2005as}
Pankaj Jain and John~P. Ralston.
\newblock {Supersymmetry and the {Lorentz} fine tuning problem}.
\newblock {\em Phys. Lett.}, B621:213--218, 2005, hep-ph/0502106.

\bibitem{ferrari:2005ng}
A.~F. Ferrari, H.~O. Girotti, and M.~Gomes.
\newblock {Lorentz symmetry breaking in the noncommutative Wess-Zumino model:
  One loop corrections}.
\newblock {\em Phys. Rev.}, D73:047703, 2006, hep-th/0510108.

\bibitem{grootnibbelink:2004za}
Stefan~Groot Nibbelink and Maxim Pospelov.
\newblock {Lorentz violation in supersymmetric field theories}.
\newblock {\em Phys. Rev. Lett.}, 94:081601, 2005, hep-ph/0404271.

\bibitem{bolokhov:2005cj}
Pavel~A. Bolokhov, Stefan~Groot Nibbelink, and Maxim Pospelov.
\newblock {Lorentz violating supersymmetric quantum electrodynamics}.
\newblock {\em Phys. Rev.}, D72:015013, 2005, hep-ph/0505029.

\bibitem{kosinski:1994vc}
P.~Kosinski, J.~Lukierski, P.~Maslanka, and J.~Sobczyk.
\newblock {Quantum deformation of the Poincare supergroup and kappa deformed
  superspace}.
\newblock {\em J. Phys.}, A27:6827--6838, 1994, hep-th/9405076.

\bibitem{kobayashi:2004ep}
Yoshishige Kobayashi and Shin Sasaki.
\newblock {Lorentz invariant and supersymmetric interpretation of
  noncommutative quantum field theory}.
\newblock {\em Int. J. Mod. Phys.}, A20:7175--7188, 2005, hep-th/0410164.

\bibitem{Irisawa:2006xx}
Manabu Irisawa, Yoshishige Kobayashi, and Shin Sasaki.
\newblock {Drinfel'd twisted superconformal algebra and structure of unbroken
  symmetries}.
\newblock {\em Prog.Theor.Phys.}, 118:83--96, 2007, hep-th/0606207.

\bibitem{Dimitrijevic:2009mt}
Marija Dimitrijevic and Voja Radovanovic.
\newblock {D-deformed Wess-Zumino model and its renormalizability properties}.
\newblock {\em JHEP}, 0904:108, 2009, 0902.1864.

\bibitem{Dimitrijevic:2011zg}
Marija Dimitrijevic, Biljana Nikolic, and Voja Radovanovic.
\newblock {Twisted SUSY: Twisted symmetry versus renormalizability}.
\newblock {\em Phys.Rev.}, D83:065010, 2011, 1101.5023.

\bibitem{Palechor:2016rkq}
C.~Palechor, A.~F. Ferrari, and A.~G. Quinto.
\newblock {Twisted Supersymmetry in a Deformed Wess-Zumino Model in (2+1)
  Dimensions}.
\newblock {\em JHEP}, 01:049, 2017, 1609.09863.

\bibitem{Berger:2001rm}
M.S. Berger and V.~Alan Kostelecky.
\newblock {Supersymmetry and Lorentz violation}.
\newblock {\em Phys.Rev.}, D65:091701, 2002, hep-th/0112243.

\bibitem{Farias:2012ed}
C.F. Farias, A.C. Lehum, J.R. Nascimento, and A.~Yu. Petrov.
\newblock {On the superfield supersymmetric aether-like Lorentz-breaking
  models}.
\newblock {\em Phys.Rev.}, D86:065035, 2012, 1206.4508.

\bibitem{Lehum:2013pca}
A.~C. Lehum, J.~R. Nascimento, A.~{\relax Yu}. Petrov, and A.~J. da~Silva.
\newblock {Supergauge theories in aether superspace}.
\newblock {\em Phys. Rev.}, D88:045022, 2013, 1305.1812.

\bibitem{Deser:1981wh}
Stanley Deser, R.~Jackiw, and S.~Templeton.
\newblock {Topologically massive gauge theories}.
\newblock {\em Ann. Phys.}, 140:372--411, 1982.

\bibitem{Schwinger:1962tp}
Julian~S. Schwinger.
\newblock {Gauge Invariance and Mass. 2.}
\newblock {\em Phys. Rev.}, 128:2425--2429, 1962.

\bibitem{abdalla2D}
M.~C.~B. Abdalla~E., Abdalla and K.~D. Rothe.
\newblock {\em {Nonperturbative methods in two-dimensional quantum field
  theory}}.
\newblock World Scientific, 1991.
\newblock 728 p.

\bibitem{Dudal:2006ip}
D.~Dudal, J.~A. Gracey, R.~F. Sobreiro, S.~P. Sorella, and H.~Verschelde.
\newblock {UV finiteness of 3D Yang-Mills theories with a regulating mass in
  the Landau gauge}.
\newblock {\em Phys. Rev.}, D75:061701, 2007, hep-th/0612024.

\bibitem{ferrari:2007mh}
A.~F. Ferrari, M.~Gomes, A.~C. Lehum, A.~Yu. Petrov, and A.~J. da~Silva.
\newblock {Perturbative finiteness of the three-dimensional Susy QED to all
  orders}.
\newblock {\em Phys. Rev.}, D77:065005, 2008, 0709.3501.

\bibitem{ferrari:2007vv}
A.~F. Ferrari et~al.
\newblock {On the finiteness of the noncommutative supersymmetric
  Maxwell-Chern-Simons theory}.
\newblock {\em Phys. Rev.}, D77:025002, 2008, 0708.1002.

\bibitem{Lehum:2015dqr}
A.~C. Lehum.
\newblock {Lorentz-violating quantum electrodynamics in two-dimensional
  aether-superspace}.
\newblock {\em EPL}, 112(5):51001, 2015, 1511.05918.

\bibitem{Borges:2013eda}
L.H.C. Borges, A.G. Dias, A.F. Ferrari, J.R. Nascimento, and A.~Yu. Petrov.
\newblock {Generation of Axion-Like Couplings via Quantum Corrections in a
  Lorentz Violating Background}.
\newblock {\em Phys.Rev.}, D89:045005, 2014, 1304.5484.

\bibitem{Arefeva:1980ms}
I.~{\relax Ya}. Arefeva and S.~I. Azakov.
\newblock {RENORMALIZATION AND PHASE TRANSITION IN THE QUANTUM CP**(n-1) MODEL
  (D = 2, 3)}.
\newblock {\em Nucl. Phys.}, B162:298--310, 1980.

\bibitem{DAdda:1978vbw}
A.~D'Adda, M.~Luscher, and P.~Di~Vecchia.
\newblock {A 1/n Expandable Series of Nonlinear Sigma Models with Instantons}.
\newblock {\em Nucl. Phys.}, B146:63--76, 1978.

\bibitem{DAdda:1978dle}
A.~D'Adda, P.~Di~Vecchia, and M.~Luscher.
\newblock {Confinement and Chiral Symmetry Breaking in CP**n-1 Models with
  Quarks}.
\newblock {\em Nucl. Phys.}, B152:125--144, 1979.

\bibitem{Abdalla:1990qf}
E.~Abdalla and F.~M. De~Carvalho~Filho.
\newblock {The Chern-Simons term and the dynamics of the CP(n-1) model in
  three-dimensions}.
\newblock {\em Int. J. Mod. Phys.}, A7:619--630, 1992.

\bibitem{asano:2003ix}
E.~A. Asano, A.~G. Rodrigues, M.~Gomes, and A.~J. da~Silva.
\newblock {The {(2+1)D} noncommutative {CP}({N}-1) model}.
\newblock {\em Phys. Rev.}, D69:065012, 2004, hep-th/0307114.

\bibitem{asano:2004vq}
E.~A. Asano, H.~O. Girotti, M.~Gomes, A.~Yu. Petrov, A.~G. Rodrigues, and A.~J.
  da~Silva.
\newblock {The coupling of fermions to the three-dimensional noncommutative
  {CP(N-1)} model: Minimal and supersymmetric extensions}.
\newblock {\em Phys. Rev.}, D69:105012, 2004, hep-th/0402013.

\bibitem{ferrari:2006xx}
A.~F. Ferrari, A.~C. Lehum, A.~J. da~Silva, and F.~Teixeira.
\newblock {The Supersymmetric (2+1){D} Noncommutative ${CP}^{(N-1)}$ Model in
  the Fundamental Representation}.
\newblock {\em J. Phys.}, A40:7803--7818, 2007, hep-th/0612223.

\bibitem{Lee:2000ey}
Bum-Hoon Lee, Ki-Myeong Lee, and Hyun~Seok Yang.
\newblock {The CP(n) model on noncommutative plane}.
\newblock {\em Phys. Lett.}, B498:277--284, 2001, hep-th/0007140.

\bibitem{Furuta:2002nv}
Ko~Furuta, Takeo Inami, Hiroaki Nakajima, and Masayoshi Yamamoto.
\newblock {NonBPS solutions of the noncommutative CP**1 model in
  (2+1)-dimensions}.
\newblock {\em JHEP}, 08:009, 2002, hep-th/0207166.

\bibitem{Furuta:2002ty}
Ko~Furuta, Takeo Inami, Hiroaki Nakajima, and Masayoshi Yamamoto.
\newblock {Low-energy dynamics of noncommutative CP**1 solitons in 2+1
  -dimensions}.
\newblock {\em Phys. Lett.}, B537:165--172, 2002, hep-th/0203125.

\bibitem{Foda:2002nt}
O.~Foda, I.~Jack, and D.~R.~T. Jones.
\newblock {General classical solutions in the noncommutative CP**(N-1) model}.
\newblock {\em Phys. Lett.}, B547:79--84, 2002, hep-th/0209111.

\bibitem{Otsu:2003fq}
Hideharu Otsu, Toshiro Sato, Hitoshi Ikemori, and Shinsaku Kitakado.
\newblock {New BPS solitons in (2+1)-dimensional noncommutative CP1 model}.
\newblock {\em JHEP}, 07:054, 2003, hep-th/0303090.

\bibitem{Ghosh:2003ka}
Subir Ghosh.
\newblock {Energy crisis or a new soliton in the noncommutative CP(1) model?}
\newblock {\em Nucl. Phys.}, B670:359--372, 2003, hep-th/0306045.

\bibitem{Otsu:2004fz}
Hideharu Otsu, Toshiro Sato, Hitoshi Ikemori, and Shinsaku Kitakado.
\newblock {Lost equivalence of nonlinear sigma and CP**1 models on
  noncommutative space}.
\newblock {\em JHEP}, 06:006, 2004, hep-th/0404140.

\bibitem{Gomes:2011aa}
M.~Gomes, J.~R. Nascimento, A.~{\relax Yu}. Petrov, and A.~J. da~Silva.
\newblock {All-loop finiteness of the two-dimensional noncommutative
  supersymmetric gauge theory}.
\newblock {\em EPL}, 98(2):21002, 2012, 1112.2105.

\bibitem{Gates:1983nr}
S.~J. Gates, Marcus~T. Grisaru, M.~Rocek, and W.~Siegel.
\newblock {Superspace, or one thousand and one lessons in supersymmetry}.
\newblock {\em Front. Phys.}, 58:1--548, 1983, hep-th/0108200.

\bibitem{Carroll:2008pk}
Sean~M. Carroll and Heywood Tam.
\newblock {Aether Compactification}.
\newblock {\em Phys. Rev.}, D78:044047, 2008, 0802.0521.

\bibitem{Gomes:2009ch}
M.~Gomes, J.R. Nascimento, A.~Yu. Petrov, and A.J. da~Silva.
\newblock {On the aether-like Lorentz-breaking actions}.
\newblock {\em Phys.Rev.}, D81:045018, 2010, 0911.3548.

\bibitem{coleman1988aspects}
S.~Coleman.
\newblock {\em Aspects of Symmetry: Selected Erice Lectures}.
\newblock Cambridge University Press, 1988.

\bibitem{ferrarisusymath}
A.~F. Ferrari.
\newblock {SusyMath: {A} Mathematica package for quantum superfield
  calculations}.
\newblock {\em Comput. Phys. Commun.}, 176:334--346, 2007,
  {http://fma.if.usp.br/$\sim$alysson/SusyMath}.

\bibitem{Gates:1977hb}
S.~James Gates, Jr.
\newblock {Supersymmetry and Yang-Mills Invariance in (1+1)-Dimensions}.
\newblock 1977.

\bibitem{Bengtsson:1983in}
Anders K.~H. Bengtsson and Ingemar Bengtsson.
\newblock {Some Properties of Supersymmetric {QED} in (1+1)-dimensions}.
\newblock {\em Nucl. Phys.}, B231:157--171, 1984.

\end{thebibliography}
\end{document}